\documentclass[sigconf]{acmart}

\usepackage{algorithmic}
\usepackage{graphicx}
\usepackage{textcomp}
\usepackage[utf8]{inputenc}
\usepackage{url}
\usepackage{hyperref}
\usepackage{pifont}
\usepackage{xspace}
\usepackage{etoolbox}
\usepackage{tcolorbox}
\usepackage{listings}
\usepackage{glossaries-extra}
\setabbreviationstyle[acronym]{long-short}
\usepackage{graphicx}
\usepackage{array}
\usepackage[english]{babel}
\usepackage{booktabs}
\usepackage{caption}
\usepackage{enumitem}
\usepackage{float}
\usepackage{graphicx}
\usepackage{ifthen}
\usepackage[utf8]{inputenc}
\usepackage{lipsum}
\usepackage{multirow}
\usepackage[group-separator={,},per-mode=symbol,binary-units=true]{siunitx}
\usepackage{subcaption}
\usepackage{tabu}
\usepackage{textcomp}
\usepackage{xcolor}
\usepackage{xspace}
\usepackage[normalem]{ulem}
\usepackage{cleveref}
\usepackage{colortbl}
\usepackage{adjustbox}

\copyrightyear{2026}
\acmYear{2026}
\setcopyright{iw3c2w3g}
\acmConference[ICSE '26]{2026 IEEE/ACM 48th International Conference on Software Engineering}{April 12--18, 2026}{Rio de Janeiro, Brazil}
\acmBooktitle{2026 IEEE/ACM 48th International Conference on Software Engineering (ICSE '26), April 12--18, 2026, Rio de Janeiro, Brazil}
\acmPrice{}
\acmDOI{10.1145/3744916.3764522}
\acmISBN{979-8-4007-2025-3/26/04}

\begin{document}

\newcommand{\assignedto}[1]{\textcolor{red}{\ding{46}~Assigned to:~#1}\\}
\newcommand{\limit}[1]{\textcolor{red}{\ding{46}~Page limit:~#1}\\}
\newcommand{\ie}{\emph{i.e.,}\xspace}
\newcommand{\eg}{\emph{e.g.,}\xspace}
\newcommand{\etc}{etc.\xspace}
\newcommand{\etal}{\emph{et~al.}\xspace} 
\newcommand{\todo}[1]{\textcolor{blue}{TODO: #1}}

\newcommand{\ma}[1]{\textcolor{red}{[MA: #1]}}
\newcommand{\im}[1]{\textcolor{red}{[IM: #1]}}
\newcommand{\kv}[1]{\textcolor{green}{[KV: #1]}}

\newcommand{\chatgpt}{ChatGPT\xspace}
\newcommand{\gpt}{GPT4\xspace}
\newcommand{\deepseek}{DeepSeek Coder\xspace}
\newcommand{\speechless}{Speechless Codellama\xspace}
\newcommand{\wizardcoder}{WizardCoder\xspace}
\newcommand{\codemillenials}{Code Millenials\xspace}

\newcommand{\se}{Software Engineering\xspace} 
\newcommand{\punchline}[1]{\noindent\textbf{#1}.\xspace}

\newcommand{\newacro}[3]{
    \newacronym{#1}{#2}{#3}
    \expandafter\def\csname #1\endcsname{\gls{#1}\xspace}
    \expandafter\def\csname #1s\endcsname{\glspl{#1}\xspace}
}

\newacro{llm}{LLM}{Large Language Model}
\newacro{rpi}{RPi}{Raspberry Pi}

\newcommand{\alternatingrowcolors}{%
    \rowcolors{2}{gray!15}{white} 
}
\newenvironment{coloredtable}{\alternatingrowcolors\noindent\begin{tabular}}
    {\end{tabular}}


\definecolor{customblue}{HTML}{45B7D1}
\definecolor{brightBackground}{HTML}{ecf8fa}
\definecolor{plotblue}{HTML}{c5e3ed}

\definecolor{large}{rgb}{1, 0.7, 0.7} 
\definecolor{medium}{rgb}{1, 0.6, 0.3} 
\definecolor{small}{rgb}{1, 0.9, 0} 

\newcommand{\marktext}[2]{\adjustbox{cframe=black,bgcolor=#1}{\strut #2}}

\tcbset{
  colback=plotblue!10,
  colframe=plotblue!80!black,
  left=1mm,
  right=1mm,
  top=1mm,
  bottom=1mm,
  title={},
  boxrule=0.5pt,
  arc=2pt
}

\definecolor{codegreen}{rgb}{0,0.6,0}
\definecolor{codegray}{rgb}{0.5,0.5,0.5}
\definecolor{codepurple}{rgb}{0.58,0,0.82}
\definecolor{backcolour}{rgb}{0.95,0.95,0.92}
\definecolor{gray50}{gray}{.5}
\definecolor{gray40}{gray}{.6}
\definecolor{gray30}{gray}{.7}
\definecolor{gray20}{gray}{.8}
\definecolor{gray10}{gray}{.9}
\definecolor{gray05}{gray}{.95}

\lstdefinestyle{pythonstyle}{
    backgroundcolor=\color{backcolour},   
    commentstyle=\color{codegreen},
    keywordstyle=\color{magenta},
    numberstyle=\tiny\color{codegray},
    stringstyle=\color{codepurple},
    basicstyle=\footnotesize\ttfamily,
    breakatwhitespace=false,         
    breaklines=true,                 
    captionpos=b,                    
    keepspaces=true,                 
    numbers=left,                    
    numbersep=5pt,                  
    showspaces=false,                
    showstringspaces=false,
    showtabs=false,                  
    tabsize=2
}

\lstnewenvironment{python}[1][]
{
    \lstset{style=pythonstyle, language=Python, #1}
}
{}

\title{Generating Energy-Efficient Code via Large-Language Models -- Where are we now? 
}

\author{Radu Apsan$^\dagger$, Vincenzo Stoico$^\dagger$, Michel Albonico$^\star$, Rudra Dhar$^\ddagger$, \\ Karthik Vaidhyanathan$^\ddagger$, Ivano Malavolta$^\dagger$}
\affiliation{%
  \institution{$^\dagger$ Vrije Universiteit Amsterdam, $^\star$ Federal University of Technology of Paraná (UTFPR), $^\ddagger$ IIIT Hyderabad}
  \country{}
}
\email{r.apsan@vu.nl, v.stoico@vu.nl, michelalbonico@utfpr.edu.br, rudra.dhar@research.iiit.ac.in}
\email{karthik.vaidhyanathan@iiit.ac.in, i.malavolta@vu.nl}







\begin{abstract}
\textit{Context}. The rise of Large Language Models (LLMs) has led to their widespread adoption in development pipelines.

\noindent\textit{Goal}. We empirically assess the energy efficiency of Python code generated by LLMs against human-written code and code developed by a Green software expert.

\noindent\textit{Method}. We test 363 solutions to 9 coding problems from the EvoEval benchmark using 6 widespread LLMs with 4 prompting techniques, and comparing them to human-developed solutions. Energy consumption is measured on three different hardware platforms: a server, a PC, and a Raspberry Pi for a total of $\approx$881h (36.7 days).

\noindent\textit{Results}. Human solutions are 16\% more energy-efficient on the server and 3\% on the Raspberry Pi, while LLMs outperform human developers by 25\% on the PC. Prompting does not consistently lead to energy savings, where the most energy-efficient prompts vary by hardware platform. 
The code developed by a Green software expert is consistently more energy-efficient by at least 17\% to 30\% against all LLMs on all hardware platforms.

\noindent\textit{Conclusions}. Even though LLMs exhibit relatively good code generation capabilities, no LLM-generated code was more energy-efficient than that of an experienced Green software developer, suggesting that as of today there is still a great need of human expertise for developing energy-efficient Python code.
\end{abstract}


\maketitle

\vspace{-3mm}
\section{Introduction}\label{s:intro}

Large Language Models (LLMs) have demonstrated remarkable capabilities in understanding natural language and generating human-like text. LLMs are supporting a wide variety of \se tasks~\cite{codeassistants}, ranging from requirements~\cite{reqllm} to design, code generation, and maintenance~\cite{zheng2025towards}. 
%
Most of LLMs-related reserch in \se regarded code generation, with the emergence of models such as CodeBERT, codeLLama, and DeepseekCoder~\cite{llms4se}. This has also resulted in the development and adoption of different AI code assistants like Github Co-Pilot and Cursor. Their ability to streamline development workflows 
has positioned them as indispensable aids for both novice and expert developers~\cite{treude2025generativeaiempiricalsoftware,so_ai,codeassistants}. 

The rapid integration of LLMs into \se processes and practices raises critical questions about their long-term implications, particularly on sustainability. While much focus has been given to the performance~\cite{coignion2024performance, NEURIPS2023_43e9d647}, readability~\cite{DBLP:journals/fi/Tosi24, DBLP:conf/msr/SiddiqRZS24}, and maintainability of LLM-generated code~\cite{dillmann2024evaluationmaintain,wadhwa2024core,arun2025llms}, its energy efficiency 
 is being studied only recently~\cite{cursaru2024controlled,10765804,cappendijk2024generating}. This oversight is surprising, given the growing importance and relevance of Green AI~\cite{verdecchia2023systematic} and software sustainability in general~\cite{calero2021introduction,lago2015framing}. 
As LLMs automate more aspects of code generation, their role in propagating or mitigating such inefficiencies becomes a urgent concern.  

This work is an empirical study comparing the energy efficiency of code generated by state-of-the-art LLMs against human-written code. 
Specifically, we systematically select six of the top \llms in one of the most recent coding benchmarks for \llms (EvoEval~\cite{evoeval}): GPT4~\cite{openai2024gpt4technicalreport}, ChatGPT~\cite{chatgpt}, DeepSeek Coder 33B~\cite{zhu2024deepseek}, Speechless Codellama 34B~\cite{codellama}, Code Millenials 34B~\cite{codemillenials}, and WizardCoder 33B~\cite{luo2023wizardcoder}. 
Further, we generate Python solutions for nine \textit{difficult} coding problems in EvoEval from two different types of developers (average and Green software expert) and the 6 considered LLMs through 4 different prompting techniques. A systematic approach is followed to curate guidelines from existing research literature to develop prompts for improving the energy efficiency of the generated code. We then systematically measure the energy consumption of the generated Python code by executing it on three types of hardware devices: Server, Personal Computer (PC), and \rpi.
In total, we collect $\approx$4.6 billion energy measures for $\approx$881h (36.7 days) of sheer execution time.

Our \textbf{results} indicate that while code generated via a base prompt outperforms developer-implemented code on the PC by 25\%, the average developer outperforms \llms on the server and \rpi by 16\% and 3\%, respectively. 
There is no `most energy-efficient' prompt, but having prompts tailored around energy efficiency did demonstrate an impact on improving energy efficiency. Further, code developed by
a Green software expert consistently outperforms all other implementations across devices by at least 17\%. This result is particularly interesting since it provides evidence about the need for carrying out more research on improving the reasoning capabilities of LLMs in terms of generating energy efficient (Python) code.

The \textbf{main contributions} of this study are: (i) an extensive empirical study on the energy efficiency of Python source code generated by six different \llms, under several energy-aware prompting techniques, and one non-energy-aware prompt, 
(ii) a set of 28 guidelines for developing energy-efficient Python code derived by conducting a quasi-systematic literature review,
(iii) energy-efficient implementations for four selected coding problems developed by a Green software expert,
(iv) an in-depth discussion on the implications of our results and recommendations for researchers and practitioners, 
(v) a complete replication package~\cite{replicationpackage} containing all supporting artifacts and raw data for independent verification of the obtained results. 
This study is the first empirical investigation on the energy-efficiency of LLM-generated Python code with this size and with an explicit assessment against code that
has been \textit{intentionally} developed to be energy-efficient by an expert in the field.


The \textbf{target audience} of this study includes:
(i) \textit{developers} willing to use \llms for generating (energy-efficient) Python code, 
(ii) \textit{\llms vendors} willing to improve the capabilities of their \llms in terms of generating energy-efficient code, and 
(iii) \textit{software engineering researchers} who can use the results of this study and our suggested research directions as inspiration for their future research.

\vspace{-2mm}
\section{Background}\label{s:bg}





\noindent\textbf{\llms for code generation}.
The field of AI has witnessed a revolution with the advent of transformer-based \llms~\cite{vaswani2017attention}. These advanced neural network architectures have dramatically reshaped numerous domains. One such domain is code generation, where \llms are also called ``coding assistants''~\cite{10.1145/3581641.3584037}. 
The ability of \llms to understand and generate code has sparked significant interest among researchers and practitioners alike, with many tech players heavily investing on \llm-based coding assistants -- \eg Copilot by GitHub. 
%
Owing to the reasonably accurate code generation capabilities of \llms, coding assistants are increasingly becoming a vital component of the software development pipeline~\cite{10.1145/3586030, 10.1145/3597503.3608128, codeassistants}. 

A critical aspect for the quality of \llm-generated code is \textbf{prompting}.
A prompt is the input provided to the \llm to get a response and it is typically used to enforce rules and ensure specific characteristics of the output~\cite{white2023promptpatterncatalogenhance}. 
Prompt engineering is the process of creating prompts to get the most accurate, relevant, and useful outputs. Since \llms are highly sensitive to the phrasing and structure of their inputs, creating effective prompts is \textit{essential} to achieve the desired results.
Two widely-used prompting techniques are (i) \textit{zero-shot prompting}, where the model is expected to generate a response based solely on its pre-existing knowledge~\cite{NEURIPS2022_8bb0d291} and (ii) \textit{few-shot prompting}, which provides the model with a few examples to help the model understand the pattern or context of the request~\cite{NIPS2017_cb8da676}. 

\noindent\textbf{Green IT}.
The increasing reliance on IT technologies in our society has led to a sharp increase in energy consumption attributed to digital infrastructure~\cite{mytton2022sources}.

Energy consumption in IT systems is influenced by many factors, \eg hardware utilization, software design, and deployment strategies~\cite{guldner2024development}.
Energy consumption ($E$, in Joules) is a function of power ($P$, in Watts) over a given time period ($t$, in seconds): $E = \int_{t_0}^{t_n} P(t) dt$, 
%
%
where $t_0$ and $t_n$ are the start and end of measurement, respectively, and $P(t)$ is the instant power drawn by the system. 
Factors such as CPU utilization~\cite{10.5555/2001180.2001183}, memory access patterns~\cite{10.1145/3136014.3136031}, and software dependencies~\cite{EASE_2024, numpyvspytorch, reya2023greenpy} can increase power usage, and thus impact the overall energy usage~\cite{castor2024estimatingenergyfootprintsoftware}. 
%
%
%
Research has shown that even programming choices like selecting certain Java collections~\cite{oliveira2021improving} or Python libraries~\cite{EASE_2024}, can significantly affect energy use. 
\textbf{Green coding} focuses on minimizing energy consumption through efficient coding practices~\cite{junger2024potentialsgreencoding}. 
Green coding strategies  
include optimized algorithms, energy-efficient software design, energy-aware programming, \etc~\cite{junger2024potentialsgreencoding}. As energy consumption in IT continues to rise, integrating these principles into \se -- including AI-driven code generation -- becomes critical.




\section{Study Design}\label{s:design}

We designed and conducted this study according to established guidelines for empirical software engineering~\cite{wohlin2012experimentation,shull2007guide} and energy efficiency assessment \cite{guldner2024development,mancebo2021process}.
According to the template by Basili \etal~\cite{basili1988tame}, the \textbf{goal} of this study is to 
analyse \textit{LLM-generated code} 
for the purpose of \textit{evaluating} 
its \textit{energy efficiency} 
from the point of view of \textit{software developers, LLM vendors, and Software Engineering researchers} 
in the context of \textit{Python software development}.
This goal is achieved by answering the following research questions.

\noindent \textbf{$\mathbf{RQ_0}$ -- How does the energy efficiency of LLM-generated code vary when using different LLMs?} 
This RQ serves for setting the context of the experiment and check if different LLMs generate code with significant differences in terms of energy efficiency.  

\noindent \textbf{$\mathbf{RQ_1}$ -- What is the difference between \underline{human code} and LLM-generated code in terms of energy efficiency?}  
This RQ compares the energy efficiency of human-developed Python code and code generated using a generic prompt. It helps determine how well LLMs generate efficient code without specific prompt interventions. 

\noindent \textbf{$\mathbf{RQ_2}$ -- What is the impact of \underline{prompt engineering techniques} on the energy efficiency of LLM-generated code?} 
This RQ provides evidence about whether prompt engineering affects the energy efficiency of LLM-generated Python code. We expect to discover the characteristics of
the prompts or the prompting strategies leading to more
energy-efficient Python code, thus providing actionable insights for developers willing to generate greener Python code without retraining or fine tuning the used LLMs. 

\noindent \textbf{$\mathbf{RQ_3}$ -- What is the difference between \underline{code developed by a} \underline{Green software expert} and LLM-generated code in terms of energy efficiency?} 
This RQ assesses how code developed by a Green software expert compares to LLM-generated code.
Previous research (see \Cref{sec:related}) shows inconsistent results when about whether human-developed code is more efficient than LLM-generated one. However, no study compared how much it is different from code that has been \textit{intentionally} developed to be energy-efficient by an expert in the field; this code can be seen as an empirically-defined lower bound for the energy usage of Python code. 

\begin{figure}[b]
    \centering
    \includegraphics[width=1\columnwidth]{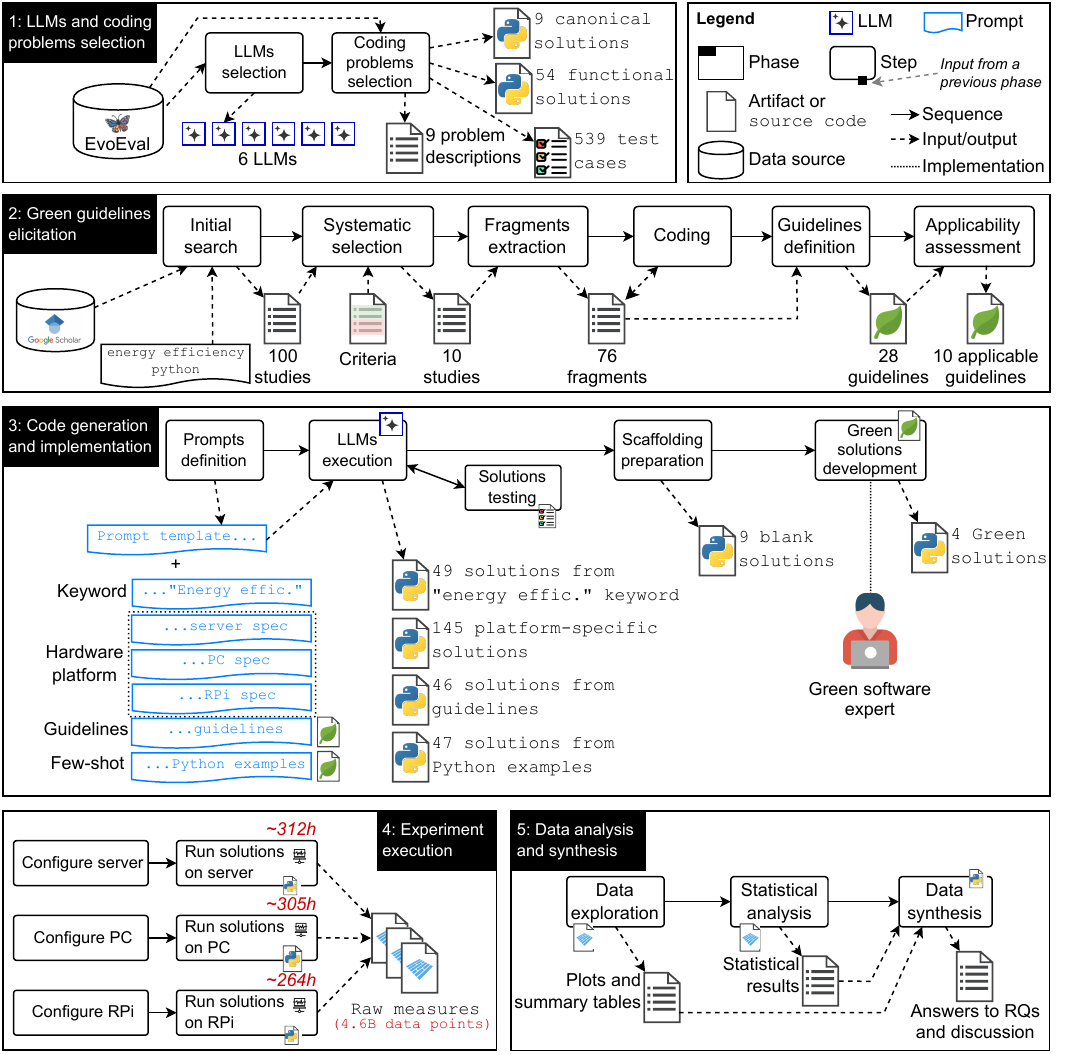}
    \vspace{-7mm}
    \caption{Overview of study.}
    \label{fig:study_design}
\end{figure}

\Cref{fig:study_design} shows the main phases of this study; each phase is described in the remainder of this section. 
To ensure independent verification, we provide a complete \textbf{replication package} \cite{replicationpackage} containing the source code, intermediate artifacts, raw and aggregated measures, and documentation related to each phase of the study.

\vspace{-1mm}
\subsection{LLMs and coding problems selection}\label{sec:phase1}

\subsubsection{LLMs selection} \label{s:design:llms}
For this study, we select the top 6 accessible, unique \llms in the \texttt{difficult} category of the EvoEval leaderboard~\cite{evoevalleaderboard}, as of July 2024 (see \Cref{tab:llm_overview}). EvoEval builds on the well-known HumanEval benchmark by OpenAI and includes 828 coding problems across different domains~\cite{evoeval}. We choose EvoEval since it (i) is one of the most recent benchmarks for evaluating \llms trained for code generation, (ii) contains human-curated ground truth solutions, (iii) includes multiple test cases for each coding problem, and (iv) includes evaluation scripts already usable in our pipeline~\cite{replicationpackage}. 
In the context of this study, an \llm is (i) \textit{accessible} if it is either open-source or available through APIs and (ii) \textit{unique} if it uses a different base model compared to other already-chosen \llms. Accessibility and uniqueness of the selected \llms facilitate the replicability and external validity of the study. 

\vspace{-3mm}
\begin{table}[htbp]
\centering
\footnotesize
\caption{\llms used in this study (* = unofficial information).}
\vspace{-4mm}
\label{tab:llm_overview}
\begin{tabular}{lrccr}
\toprule
\textbf{LLM} & \textbf{Parameters} & \textbf{Open} & \textbf{EvoEval rank} \\ \midrule
\href{https://platform.openai.com/docs/models#gpt-3-5-turbo}{ChatGPT} & 175B* & No & 6 \\ \rowcolor{gray10}
\href{https://platform.openai.com/docs/models/gpt-4}{GPT4} & 1.76T* & No & 2 \\
\href{https://huggingface.co/deepseek-ai/deepseek-coder-33b-instruct}{DeepSeek Coder Instruct} & 33B & \checkmark & 4 \\ \rowcolor{gray10}
\href{https://huggingface.co/uukuguy/speechless-codellama-34b-v2.0}{Speechless Codellama v2.0} & 34B & \checkmark & 10 \\
\href{https://huggingface.co/budecosystem/code-millenials-34b}{CodeMillenials} & 34B & \checkmark & 9 \\ \rowcolor{gray10}
\href{https://huggingface.co/WizardLMTeam/WizardCoder-33B-V1.1}{WizardCoder v1.1} & 33B & \checkmark & 5 \\
\bottomrule
\end{tabular}%
\end{table}
\vspace{-6mm}

\subsubsection{Coding problems selection} \label{s:design:coding_problems}

We start from the 100 coding problems in the \texttt{Difficult} category in EvoEval since we aim to generate Python code with higher computational complexity and additional requirements compared to the other categories in EvoEval~\cite{evoeval}. Since our goal was to evaluate the energy efficiency of the generated code given that effectiveness (correctness) was already established, we selected those problems that are correctly solved by all six selected \llms (\ie the solutions generated by all \llms pass all tests). 
This selection step leads to the following 9 coding problems: 
 (1) \texttt{weighted\_mean\_absolute\_deviation} (EvoEval ID: 4),
 (2) \texttt{count\_distinct\_characters\_substrings} (EvoEval ID: 16),
 (3) \texttt{below\_above\_threshold} (EvoEval ID: 52),
 (4) \texttt{add\_elements} (EvoEval ID: 53),
 (5) \texttt{correct\_bracketing\_advanced} (EvoEval ID: 61),
 (6) \texttt{customFibFib} (EvoEval ID: 63),
 (7) \texttt{advancedDigitSum} (EvoEval ID: 66),
 (8) \texttt{decimal\_to\_binary} (EvoEval ID: 79),
 (9) \texttt{next\_} \texttt{smallest\_and\_largest} (EvoEval ID: 90).

The selected coding problems are about tasks involving string manipulation, recursion, numerical calculations, \etc In this step we also collect the 9 \textit{canonical solutions} provided in the EvoEval dataset; those solutions have 3 to 22 lines of code and contain different programming constructs, such as (nested) loops, recursion, variable assignments, conditionals, regular expressions. From the EvoEval dataset we also collect (i) the 54 \textit{functional solutions} generated for each coding problem by the 6 LLMs selected in the previous step -- these solutions are particularly useful for answering $RQ_1$ since they have been curated by third parties (thus, no internal threat to validity for our study) and are curated by humans -- and (ii) 539 test cases (min: 20, max: 86, median: 62), which we use for assessing the correctness of the coding solutions generated in Phase 3.

\subsection{Green guidelines elicitation}\label{sec:phase2}
In this phase we systematically elicit guidelines for energy-efficient Python code. 
The guidelines elicitation process is intentionally designed to produce guidelines that are \textit{broadly applicable to any Python project}, regardless of the specific coding problems considered in this study. In this study, the guidelines are used in Phase 3 for (i) crafting prompts and (ii) manually developing Green solutions. Inspired by systematic literature reviews~\cite{kitchenham2010systematic}, we apply the following six steps.

\vspace{-2mm}
    \subsubsection{Initial search} 
    We collect the first 100 results from a query on Google Scholar, which is currently considered one of the most comprehensive academic search engines publicly available~\cite{gusenbauer2019google}.  

\vspace{-2mm}
    \subsubsection{Systematic selection}
    We manually analyse each of the 100 potentially-relevant studies and rigorously assess whether it is about energy efficiency of software, peer-reviewed, and in English. Following the guidelines for systematic literature review for software engineering~\cite{kitchenham2007guidelines}, we define a a priori a set of inclusion and exclusion criteria (see replication package).
    To avoid bias, this analysis is carried out in parallel by two independent researchers, which achieved a 94\% level of agreement, with Cohen Kappa = 0.67 (\textit{substantial}) \cite{mchugh2012interrater}. After discussing the 6 studies with conflicts, we obtain a final set of 10 relevant studies.  

    \vspace{-2mm}
    \subsubsection{Fragments extraction}
    Three researchers independently extract from the 10 studies verbatim text fragments containing indications about how to implement energy-efficient Python code.  

    \vspace{-2mm}
    \subsubsection{Coding} 
    Three researchers independently apply a process inspired by the (open) card sorting technique~\cite{spencer2009card} in two phases: 
    (i) we code each fragment with its representative information (\eg \texttt{use compiled languages for critical code sections}) and (ii) we group the extracted codes into meaningful families with a descriptive title (\eg \texttt{threading}, \texttt{parallel computation}).

    \vspace{-2mm}
    \subsubsection{Guidelines definition} 
    Similar codes are consolidated into broader themes, which are then summarized into 28  guidelines (see Table~\ref{tab:guidelines}). Each guideline is defined as a one-sentence recommendation that is widely applicable across different software projects/organizations. Each guideline is also categorized according to 7 families, namely: Code optimization, Multithreading, Native code, Function calls, Object-orientation, Network, and Other. 

    \vspace{-2mm}
    \subsubsection{Applicability assessment}
    Two researchers assess whether each guideline is directly applicable to at least one of the 9 coding problems selected in Phase 1, leading to the final set of 10 Green guidelines we will use in Phase 3. 

    \begin{table*}[t]
    \caption{Guidelines for energy-efficient Python code. The asterisk (*) marks the guidelines that were selected for this study.}
    \vspace{-3mm}
    \label{tab:guidelines}
    \footnotesize
    \begin{tabular}{l|p{12.6cm}|l|p{1.3cm}}
    \toprule
    \textbf{ID} & \textbf{Guideline} & \textbf{Family} & \textbf{Provenance} \\ \hline

    G1* & When there are multiple occurrences of the same expression, assign it to a variable and use the variable. & Code optimization & \cite{3EZYN6BB} \\ \rowcolor{gray10}
    G2* & Avoid redundant operations in sorting already sorted or semi-sorted collections & Code optimization & \cite{AJRGBTUR} \\ 
    G3* & Use loop optimization techniques (\eg loop unrolling, loop unswitching, early termination) to improve performance, such as storing loop end condition in a variable & Code optimization & \cite{S2ZTE2XT} \cite{3EZYN6BB} \cite{AJRGBTUR} \\ \rowcolor{gray10}
    G4* & Use short-circuit versions of logical operators where the second argument of a logical expression is evaluated only if the first argument is insufficient to determine the value of the expression. & Code optimization & \cite{3EZYN6BB} \\ 
    G5 & When high-precision computation is not needed, approximate computations & Code optimization & \cite{3EZYN6BB} \\ \rowcolor{gray10}
    G6 & Avoid storing or saving data that has already been computed, saved, or is unnecessary. For example, don't recompute data that is available in a local variable, or avoid having variables only for debugging purposes. & Code optimization & \cite{S2ZTE2XT} \cite{3EZYN6BB} \\ 
    G7 & Use bulk operations to reduce overhead from handling individual tasks, \eg generating a batch of integers at once rather than one by one. & Code optimization & \cite{3EZYN6BB} \cite{LV5URM94} \\ \rowcolor{gray10}
    G8 & Delegate parallelizable computations to external modules such as ctypes to avoid bottlenecks caused by Python's Global Interpreter Lock. & Multithreading & \cite{H7TKU62K} \cite{34T73VA3} \cite{3EZYN6BB} \\ 
    G9 & When possible, reduce data dependencies between iterations of the same code section to make the code parallelizable. & Multithreading & \cite{TE34LYI6} \cite{S2ZTE2XT} \cite{AEVEQL99} \\ \rowcolor{gray10}
    G10 & For parallelized code, use multithreading frameworks, such as \texttt{PyOpenGL} and \texttt{mpi4py}. & Multithreading & \cite{S2ZTE2XT} \cite{AEVEQL99} \cite{AJRGBTUR} \\ 
    G11 & For multithreading, use thread shuffling to group slow threads on the same core through migration, improving execution efficiency. & Multithreading & \cite{3EZYN6BB} \\ \rowcolor{gray10}
    G12 & For multithreading, enable work-stealing dynamic scheduling to utilize idle processor queues and accelerate execution. & Multithreading & \cite{3EZYN6BB} \\ 
    G13 & Put processes or threads on sleep state if they are waiting for I/O operations or they are no longer active. & Multithreading & \cite{3EZYN6BB} \\ \rowcolor{gray10}
    G14 & Cache frequently used data shared by processes or nodes. & Multithreading & \cite{AEVEQL99} \\ 
    G15* & Write the energy- and performance-critical code sections of a Python program using a compiled language (\eg C, C++, Rust) using SWIG wrappers, \texttt{Numba}, or \texttt{Cython}. & Native code & \cite{TE34LYI6} \cite{H7TKU62K} \cite{S2ZTE2XT} \cite{LV5URM94} \cite{EASE_2024} \cite{AJRGBTUR} \\ \rowcolor{gray10}
    G16 & Use compiler optimization flags to enhance the performance of compiled language code in Python. & Native code & \cite{S2ZTE2XT} \\ 
    G17 & Avoid repeatedly calling heavy functions (\eg calls to the NumPy library) within loops.  & Native code & \cite{AJRGBTUR} \\ \rowcolor{gray10}
    G18 & Use memoization to avoid repeated execution of expensive pure functions. & Function calls & \cite{3EZYN6BB} \cite{AJRGBTUR} \\ 
    G19 & Reduce function calls to minimize overhead; use static methods and inline code to avoid lookup delays. & Function calls & \cite{3EZYN6BB} \\ \rowcolor{gray10}
    G20* & Reduce the number of created objects to minimize the memory used for creating them. & Object-orientation & \cite{3EZYN6BB} \\ 
    G21* & Use energy-efficient design patterns like Flyweight, Mediator, and Proxy, and avoid energy-intensive patterns like Decorator and Abstract Factory. & Object-orientation & \cite{3EZYN6BB} \\ \rowcolor{gray10}
    G22 & Reduce network utilization by compressing data before transmitting it. & Network & \cite{3EZYN6BB} \\ 
    G23 & Perform complex operations using more powerful hardware, such as cloud-based resources, to achieve better performance. & Network & \cite{3EZYN6BB} \\ \rowcolor{gray10}
    G24* & For computationally intensive tasks, such as, matrix-matrix multiplications, eigenvalue computation and fast fourier transform, use High-Performance Computing libraries like \texttt{NumPy} and \texttt{SciPy}. & Other & \cite{TE34LYI6} \cite{S2ZTE2XT} \cite{C6FPNKTI} \cite{AEVEQL99} \cite{AJRGBTUR} \\ 
    G25 & For parallelizable complex workload, such as, matrix multiplication and pointwise evaluation, offload the workload to accelerators like GPUs. & Other & \cite{TE34LYI6} \cite{34T73VA3} \cite{3EZYN6BB} \\ \rowcolor{gray10}
    G26 & Allocate all available computational resources (\eg CPU cores) to a process to improve execution time and energy efficiency. & Other & \cite{AEVEQL99} \cite{EASE_2024} \\ 
    G27* & Replace native Python data structures (\eg \texttt{List}) with high-performance alternatives like HDF5 for better parallelization and efficiency. & Other & \cite{AEVEQL99} \cite{3EZYN6BB} \\ \rowcolor{gray10} 
    G28* & Minimize memory accesses to improve performance and energy efficiency, \eg by aggregating operations to reduce I/O calls. & Other & \cite{3EZYN6BB} \\

    \bottomrule  
    \end{tabular}
\end{table*}



\subsection{Code generation and implementation}\label{sec:phase3}
In this phase we obtain the Python source code of all solutions; we will run and measure it in Phase 4 (see \Cref{fig:study_design}). 

\vspace{-4mm}
\subsubsection{Prompts definition}
We consider the original prompt in EvoEval as our \textit{prompt template}, such prompt is used for generating  the 54 \texttt{functional solutions} described in \Cref{sec:phase1}.
Then, we consider four different variations of the prompt template: keyword, hardware platform, guideline, and few-shot.
In the \textit{keyword} prompt we append the following sentence at the end of the prompt template: ``Generate the following code for me such that this is 'energy efficient'''.  With this prompt we assess whether LLMs have intrinsic knowledge about energy efficiency and are able to generate energy efficient software without any additional information from the developer. 
%
In the \textit{hardware platform} prompts we include the technical specification of the hardware where the generated Python code will run on. With this prompt we investigate on whether LLMs are able to generate Python code that can take advantage of hardware characteristics (\eg presence of a GPU, limited memory, \etc). Since the experiment will be executed on three different machines, \ie a server, a personal computer, and a Raspberry Pi (see \Cref{sec:phase4}), we generate one prompt for each of them.  
The \textit{guidelines} prompts include the 10 guidelines elicited in Phase 2.
To keep the context provided to the LLM as focused as possible, we create a prompt specific to each of the 9 coding problems where we include only the subset of guidelines that can be directly applicable to the coding problem.
In \textit{few-shot} prompts we include examples of code optimization for energy efficiency. Each example is composed of a pair of Python snippets, where the first one is a textbook implementation of the \texttt{bubble sort} algorithm and the second one is an energy-efficient version of the same algorithm obtained by applying a Green guideline for each family elicited in Phase 2. Both implementations are validated by running an ad-hoc experiment to ensure that (i) both implementations are correct and (ii) the energy-efficient version consumes less energy than the base one.

\vspace{-2mm}
\subsubsection{LLMs execution}
In this step the 6 \llms are prompted to generate solutions for each of the 9 coding problems.
The prompting is done automatically for each of the \llms.
ChatGPT, GPT4, and DeepSeek Coder are prompted via their proprietary APIs, whereas Speechless Codellama, Code Millenials, and WizardCoder are deployed using HuggingFace inference endpoints on the AWS.
Once the \llm generates the response, it is saved separately in its raw format, then filtered for any additional text tags, and other irrelevant lines of code (\ie example use cases, or \texttt{main()} functions), and saved to a separate file from the raw response.
A visual inspection is carried out by one researcher for identifying issues regarding typos, duplicate lines of code, or wrong function names. This checking step does not change the semantics of the generated program, but tackles generic syntax mistakes.

\vspace{-2mm}
\subsubsection{Solutions testing}
In this step we test the correctness of each generated solution using the 539 test cases collected in Phase 1. 
If a solution is deemed incorrect (\ie any of the test cases fail), then we reprompt the LLM for the current coding problem. 
If a solution for a given LLM and a given prompt is not correct after 6 reprompts, the solution is excluded from the study. Refer to \Cref{fig:study_design} for the number of correct solutions for each type of prompt.


\vspace{-2mm}
\subsubsection{Scaffolding preparation}
\label{subsubsec-scaffolding}

In this phase we prepare the so-called \texttt{blank solution} for each coding problem, \ie 
the Python code containing (i) \texttt{sol}, a function that currently contains a \texttt{pass} instruction, but later will contain the source code of the generated solution, (ii) \texttt{tc}, the set of test cases for \texttt{sol}, (iii) \texttt{n}, the number of iterations in which \texttt{sol} must be executed on all entries in \texttt{tc}, (iv) a main function sequentially calling \texttt{sol} on all entries in \texttt{tc}, \texttt{n} times.

Blank solutions act as a baseline since they isolate the overhead of the Python interpreter, operating system, and scaffolding Python code from the energy consumption of the LLM-generated solutions.
These values are used as reference points, and are not subtracted from any of the energy readings throughout this study to preserve the original datapoints.

\vspace{-2mm}
\subsubsection{Green solutions development}
In this step we produce \texttt{Green solutions}, the best possible implementations of the coding problems from an energy efficiency point of view. 
The Green solutions are manually developed by a senior developer with over 7 years of experience in Green software development.
For each coding problem, the expert is given as input (i) the description of the coding problem to solve, (ii) its canonical solution (see \Cref{sec:phase1}), (iii) its test cases, and (iv) the set of guidelines for energy-efficient Python code (see \Cref{sec:phase2}), and they are tasked to make the given canonical solution as energy-efficient as possible, while passing all given test cases. The expert is tasked to develop the Green solutions by applying as much as possible their own technical expertise and by using all the inputs listed above. 
The expert produced \texttt{Green solutions} for 4 of the 9 coding problems since problems 16, 52, 53, 66, and 79 have low cyclomatic complexity with little for improving their source code. 

At the end of Phase 3 we have a total of 363 solutions, which will be executed and measured in Phase 4.

\vspace{-1mm}
\subsection{Experiment execution}\label{sec:phase4}
The goal of this phase is to execute the 363 solutions on three different hardware platforms: a server, a personal computer (PC), and a Raspberry Pi (RPi). The reason for using three different platforms is to cover a wide spectrum of hardware specifications, so to be able to generalize the obtained results across different hardware specifications, thus strengthening the external validity of the experiment. 
The hardware platform is a blocking factor in our experiment.

\vspace{-2mm}
\subsubsection{Platforms configuration}
The technical specifications of the chosen platforms are summarized in \Cref{tab:machine_specs}. 
For all machines we ensure that all non-critical processes are stopped to reduce overhead, and we set the power governor (i.e., CPUFreq) to performance. 
We install energy tracking tools on the machines: EnergiBridge on the server and PC and Monsoon Power Meter on the \rpi, along with the necessary prerequisites for running the generated code on the test devices, including any Python packages that the code imports. As the Monsoon Power Meter requires a companion machine to track energy usage on the \rpi, the Monsoon Python package is installed on companion machines along with an HTTP server that allows the measurement to be remotely controlled. We use the SAR tool on the \rpi to profile code performance. 

\vspace{-3mm}
\begin{table}[htbp]
    \caption{Technical specifications of the hardware platforms.}
    \vspace{-4mm}
    \label{tab:machine_specs}
    \footnotesize
    \begin{tabular}{l|p{1.9cm}|p{1cm}|r|p{1cm}|l}
    \toprule
        \textbf{Platform} &  \textbf{CPU} & \textbf{GPU} & \textbf{RAM} & \textbf{OS} & \textbf{Python}  \\ \hline
        Server & Intel Xeon Silver 4208 @2.10GHz & - & 384Gb & Ubuntu 20.04 & 3.8\\ \rowcolor{gray10}
        PC & Intel Core i9 @3.50 GHz & Nvidia RTX 4070 & 64Gb & Ubuntu 24.04 & 3.12 \\
        RPi & Cortex-A72 (ARM v8) @1.8GHz & - & 8Gb & Raspberry Pi OS & 3.11 \\
         
        \bottomrule
    \end{tabular}
\end{table}
\vspace{-2mm}

The experiment is orchestrated via Experiment Runner~\cite{S2_Group_Experiment_Runner}, an open-source tool for orchestrating measurement-based experiments used in several studies in the literature~\cite{S2_Group_Experiment_Runner}. 
To avoid influencing the energy readings on the hardware platforms, Experiment Runner is executed on a dedicated machine, the experiment manager, which communicates with the server, PC, and \rpi via SSH. 
%
We measure the energy efficiency of the test platforms using EnergiBridge~\cite{Sallou_EnergiBridge_Empowering_Software_2023, durán2024energyconsumptioncodesmall, roque2024unveilingenergyvampiresmethodology} for PC and server, and the Monsoon Power Monitor~\cite{monsoon} for \rpi~\cite{en12010027, schulman2011phone, 10.1145/2380445.2380502}.
The data generated by the energy measurement tools is converted to Parquet files to optimize storage use.

\vspace{-2mm}
\subsubsection{Run solutions}

In this phase we execute the 363 solutions on the server, PC, and Raspberry Pi. The input for each solution is the same as the 539 test cases. 
The execution follows two main phases:

\noindent \textbf{Phase 1:} Determining the number of iterations. In this phase, we aimed to determine $X^m_p$ , i.e., the number of iterations for all test cases of problem p on machine m, to achieve an approximate running time of 3 minutes in the second phase (to ensure sufficient energy samples). Specifically, for each problem p, we ran its 9 functional solutions, iterating 10,000 times over all its test cases, and repeated this process 30 times to capture variability. $X^m_p$ was then calculated as: (3 minutes×10000)/(mean execution time of the solution of p on m over 30 repetitions). This process ensures that the execution time for each solution variant in the main experiment would be approximately three minutes.

\noindent \textbf{Phase 2:} Measuring energy consumption. We then measured the energy consumption for all 363 solution variants across the 3 machines. On each machine m, we executed all test cases for each variant of a solution to problem p for $X^m_p$ iterations. Each of these runs was repeated 21 times on each hardware platform to (i) take into account the intrinsic fluctuations of energy measures and (ii) strengthen statistical power~\cite{guldner2024development}. 
The experiment execution took $\approx$881 hours across all machines and generated a total of $\approx$4.6B data points.
We do not subtract the energy consumption of a machine’s idle state, as this avoids introducing errors and better reflects real-world costs. Instead, we use \emph{blank solutions} (see Section~\ref{subsubsec-scaffolding}), which isolate the overhead of scaffolding code and reveal the additional energy used for computation.

    
    




\subsection{Data analysis and synthesis}\label{sec:phase5}

\subsubsection{Data exploration}
We compute summary statistics and apply data summarization and visualization techniques based on tables, histograms, boxplots, and violin plots.
For example, Table \ref{tab:descriptive_statistics} presents descriptive statistics for energy consumption, grouped by LLM and prompt type to directly address our RQs. The table is structured in two parts: the top panel aggregates metrics by LLM (averaged over all prompt types), and the bottom panel aggregates by prompt type (averaged over all LLMs). Data is aggregated in this manner because our RQs focus on the LLMs and prompts, rather than the characteristics of individual solutions. Descriptive statistics are discussed using the median, as the distribution of all our samples was found to be non-normal. A qualitative analysis of the generated source code is provided in the Discussion section.
For the sake of brevity, in this paper we report only a part of the exploration artifacts, leaving the others in the replication package~\cite{replicationpackage}.
To further ensure the accuracy of our analysis, we employ the Interquartile Range (IQR) method to remove outliers in the energy metric. 

\vspace{-2mm}
\subsubsection{Statistical analysis}
We firstly assess the distribution of the data via Q-Q plots and the Shapiro-Wilk test (for checking normality, $\alpha$ = 0.05).
We anticipate that the collected measures do not follow a normal distribution, thus, after checking other assumptions, we statistically analyse the collected energy measure using the Aligned Rank Transformation (ART) ANOVA test. ART ANOVA is a non-parametric method that allows for multiple independent variables, interactions, and repeated measures~\cite{10.1145/1978942.1978963}.
When applying the ART ANOVA test, we (i) have two main factors, namely the used LLMs (with 6 treatments, one for each LLM) and the type of solution (with 7 levels, \eg canonical, functional, ``energy efficiency'' keyword, platform, guidelines, \etc), (ii) consider the energy efficiency of each experiment run as response variable. 
Post hoc analysis is conducted via the Aligned Ranked Transform Contrasts~\cite{elkin2021aligned} and the obtained p-values are corrected via the Holm-Bonferroni procedure~\cite{holm1979simple}. 
We use Cliff’s Delta~\cite{cliff1993dominance} for \textit{effect size estimation} and interpret its results as proposed by Vargha and Delaney~\cite{vargha2000critique}.
For $RQ_3$, we statistically analyze only the subset of data about the 4 coding problems for which we have solutions developed by the Green expert.

\vspace{-2mm}
\subsubsection{Data synthesis}
We (i) explicitly answer the RQs of this study based on the results of the previous analysis steps and (ii) contextualize the answers to the RQs by elaborating on relevant findings and implications for developers, researchers, and LLM maintainers.

\begin{table*}[h!]
    \centering
    \caption{Energy Consumption (kJ) Statistics for LLMs and prompt type. Legend - SRV : Server, Std: Standard Deviation}
    \vspace{-4mm}
    \label{tab:descriptive_statistics}
    \footnotesize{
        \begin{tabular}{p{0.5cm}rrr|rrr|rrr|rrr|rrr|rrr}
        \toprule
        & \multicolumn{3}{c}{\textbf{\chatgpt}} & \multicolumn{3}{c}{\textbf{\codemillenials}} & \multicolumn{3}{c}{\textbf{\deepseek}} & \multicolumn{3}{c}{\textbf{\gpt}} & \multicolumn{3}{c}{\textbf{\speechless}} & \multicolumn{3}{c}{\textbf{\wizardcoder}} \\ \midrule
        & \multicolumn{1}{c}{SRV} & \multicolumn{1}{c}{PC} & \multicolumn{1}{c}{RPi} & \multicolumn{1}{c}{SRV} & \multicolumn{1}{c}{PC} & \multicolumn{1}{c}{RPi} & \multicolumn{1}{c}{SRV} & \multicolumn{1}{c}{PC} & \multicolumn{1}{c}{RPi} & \multicolumn{1}{c}{SRV} & \multicolumn{1}{c}{PC} & \multicolumn{1}{c}{RPi} & \multicolumn{1}{c}{SRV} & \multicolumn{1}{c}{PC} & \multicolumn{1}{c}{RPi} & \multicolumn{1}{c}{SRV} & \multicolumn{1}{c}{PC} & \multicolumn{1}{c}{RPi} \\        \midrule
        \textbf{Mean} & 5.640 & 9.100 & 0.475 & 6.408 & 8.001 & 0.418 & 6.585 & 8.441 & 0.430 & 8.862 & 9.075 & 0.477 & 4.953 & 8.383 & 0.426 & 6.764 & 8.735 & 0.396 \\
        \textbf{Min} & 3.640 & 5.640 & 0.350 & 0.644 & 0.881 & 0.060 & 2.602 & 3.820 & 0.234 & 3.527 & 5.073 & 0.324 & 2.472 & 4.055 & 0.222 & 3.570 & 4.143 & 0.250 \\
        \textbf{50\%} & 5.582 & 9.468 & 0.473 & \cellcolor{pink} 6.341 & \cellcolor{lime} 8.480 & 0.411 & 6.233 & 8.576 & 0.407 & 5.598 & \cellcolor{pink} 9.545 & \cellcolor{pink} 0.480 & \cellcolor{lime} 5.115 & 9.393 & 0.454 & 6.026 & 8.570 & \cellcolor{lime} 0.399 \\
        \textbf{Max} & 8.861 & 12.591 & 0.744 & 11.404 & 10.875 & 0.771 & 13.004 & 12.546 & 0.678 & 75.168 & 18.304 & 0.828 & 7.562 & 10.628 & 0.533 & 24.436 & 27.672 & 0.671 \\
        \textbf{Std} & 0.878 & 1.526 & 0.064 & 1.530 & 1.628 & 0.089 & 1.950 & 2.170 & 0.100 & 12.863 & 2.172 & 0.082 & 1.033 & 1.993 & 0.076 & 3.328 & 3.838 & 0.071 \\
        \textbf{CV} & 0.156 & 0.168 & 0.135 & 0.239 & 0.204 & 0.213 & 0.296 & 0.257 & 0.232 & 1.452 & 0.239 & 0.173 & 0.209 & 0.238 & 0.179 & 0.492 & 0.439 & 0.179 \\
        \bottomrule
        \toprule
        & \multicolumn{3}{c}{\textbf{Canonical}} & \multicolumn{3}{c}{\textbf{Functional}} & \multicolumn{3}{c}{\textbf{Keyword}} & \multicolumn{3}{c}{\textbf{Platform}} & \multicolumn{3}{c}{\textbf{Guideline}} & \multicolumn{3}{c}{\textbf{Few-shot}} \\
        \midrule
        \textbf{Mean} & 5.131 & 11.345 & 0.442 & 6.082 & 8.588 & 0.452 & 6.293 & 8.728 & 0.441 & 6.111 & 8.492 & 0.442 & 7.525 & 8.663 & 0.412 & 6.880 & 8.670 & 0.450 \\
        \textbf{Min} & 3.355 & 6.523 & 0.326 & 2.596 & 4.093 & 0.267 & 2.587 & 3.855 & 0.242 & 2.485 & 3.850 & 0.245 & 0.644 & 0.881 & 0.060 & 2.472 & 3.820 & 0.234 \\
        \textbf{50\%} & \cellcolor{lime} 5.105 & \cellcolor{pink} 12.396 & 0.434 & 5.785 & 8.681 & \cellcolor{pink} 0.460 & 5.829 & \cellcolor{lime} 8.579 & 0.445 & \cellcolor{pink} 5.872 & 8.838 & 0.443 & 5.833 & 9.010 & \cellcolor{lime} 0.407 & 5.416 & 8.699 & 0.441 \\
        \textbf{Max} & 7.294 & 14.161 & 0.559 & 11.301 & 12.007 & 0.671 & 24.436 & 27.672 & 0.744 & 12.059 & 12.591 & 0.723 & 75.168 & 13.540 & 0.771 & 53.096 & 18.304 & 0.828 \\
        \textbf{Std} & 0.688 & 2.216 & 0.062 & 1.384 & 1.597 & 0.063 & 2.828 & 3.229 & 0.078 & 1.504 & 1.732 & 0.084 & 10.543 & 2.206 & 0.105 & 7.431 & 2.309 & 0.095 \\
        \textbf{CV} & 0.134 & 0.195 & 0.140 & 0.228 & 0.186 & 0.140 & 0.449 & 0.370 & 0.178 & 0.246 & 0.204 & 0.191 & 1.401 & 0.255 & 0.255 & 1.080 & 0.266 & 0.211 \\
        \bottomrule
        \end{tabular}
    }
\end{table*}

\section{Results}
\label{s:results}


\subsection{Efficiency of LLM-generated code ($\mathbf{RQ_0}$)}
\label{sec:results_rq0}
\noindent \textbf{Data Exploration}. 
We observe different results in all three hardware platforms (see \Cref{tab:descriptive_statistics}), which confirms that segmenting our experiment by hardware platform was a good choice. 
On the server, the most energy efficient solutions were those generated by \speechless ($5.115$ kJ), followed by those produced by \chatgpt ($5.582$ kJ). In contrast, the solutions generated by \speechless exhibit the highest median energy consumption ($6.341$ kJ).
On the PC, \codemillenials solutions are the most energy-efficient ($8.480$ kJ), followed by those generated by \wizardcoder ($8.470$ kJ), and \gpt generated the less energy-efficient solutions ($9.545$ kJ). 
On the RPi, \wizardcoder solutions are the most energy-efficient  ($0.399$ kJ), while \gpt solutions are the less energy-efficient ($0.480$ kJ). It is worth noting the high dispersion ($CV > 1.4$) of the \gpt solutions on the server, which led us to a deeper investigation of the impact of the other factors of this study, namely the prompts.

\noindent \textbf{Statistical Analysis}. The Shapiro-Wilk test and the Q-Q plots suggest non-normal data on each hardware platform. 
Thus, we use the ART ANOVA and the Cliff's Delta test to quantify the differences between the groups. \Cref{tab:p_values_effect_size} reports the results of both tests. The ART ANOVA gives a \textit{significant difference} between the energy efficiency of the solutions generated by different LLMs \textit{on all hardware platforms} (p-value $< 2.22e-16$). The post-hoc analysis, consisting of a pairwise comparison among the LLMs, confirms the observations of our data exploration. The energy usage of generated code is significantly different across all LLMs (p-value $<1e-4$), except for a few cases; specifically, the energy efficiency of generated solutions are not significantly different for (i) \chatgpt and \gpt on PC ($0.16$) and RPi ($0.88$), (ii) \codemillenials and \deepseek on both server and RPi ($0.088$), and (iii) \codemillenials and \wizardcoder on PC ($0.107$).

\begin{figure}[t]
    \centering
    \includegraphics[width=\linewidth]{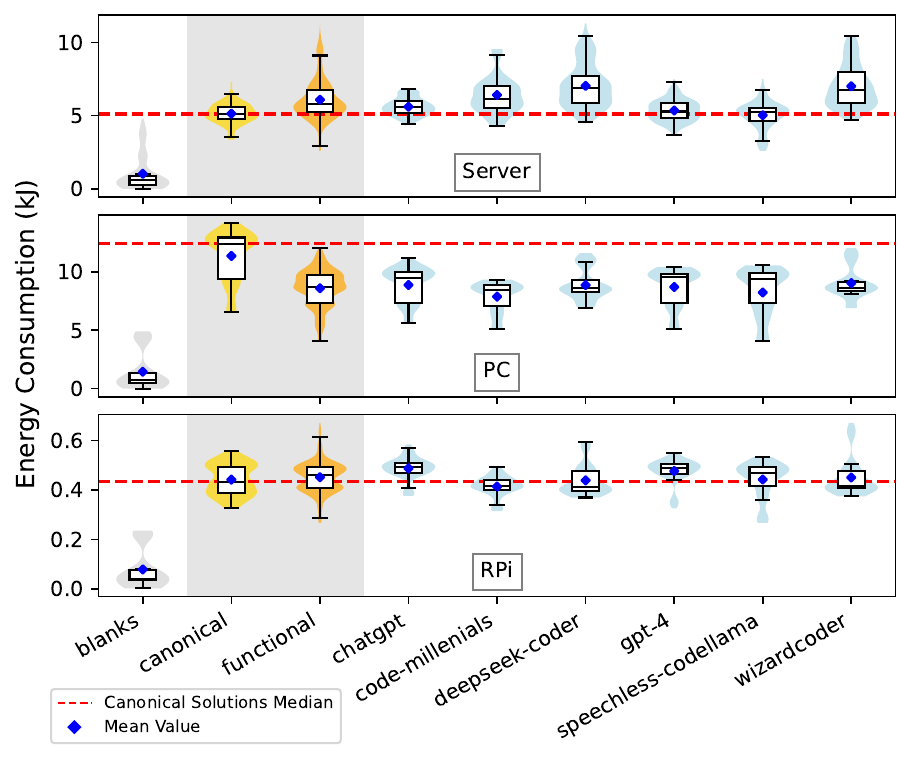}
    \vspace{-6mm}
    \caption{Energy of canonical and functional solutions.}
    \label{fig:energy_violinplot}
\end{figure}

The Cliff's delta test confirms that the solutions generated by \speechless are more energy-efficient on the server. We find a \textit{large} effect size when comparing to \speechless solutions those generated with \codemillenials ($0.637$), \deepseek ($0.548$), and \wizardcoder ($0.529$), while the effect size is medium compared to \chatgpt ($0.36$) and \gpt ($0.362$). Additionally, we notice \textit{medium} effect size contrasting \chatgpt to \codemillenials ($0.409$) solutions. The rest of the comparisons result in either \textit{small} or \textit{negligible} effect size. The effect size results \textit{medium} on the PC confronting \chatgpt to \codemillenials ($-0.457$) and \wizardcoder ($-0.39$), as well as \gpt with \codemillenials ($-0.398$) and \wizardcoder ($-0.335$). We notice a \textit{large} effect size on RPi comparing \chatgpt with \codemillenials ($-0.577$) and \wizardcoder ($-0.68$), as well as \codemillenials and \gpt ($0.527$) and the latter against \wizardcoder ($-0.634$). This result confirms the lower energy consumption of \wizardcoder solutions observed in \Cref{tab:descriptive_statistics}.

\begin{table}[htbp]
    \centering
    \caption{P-values of pairwise comparison and effect sizes.}
    \vspace{-4mm}
    \label{tab:p_values_effect_size}
    \resizebox{\columnwidth}{!}{%
       \begin{tabular}{llrrr}
            \toprule
            \multicolumn{2}{c}{\textbf{Contrast}} & 
            \multicolumn{3}{c}{\textbf{P-value (effect size)}} \\
            \midrule
            & & \multicolumn{1}{c}{Server}  & \multicolumn{1}{c}{PC}  & \multicolumn{1}{c}{RPi}  \\
            chatgpt & code-millenials & \cellcolor{medium} $\mathbf{<1e-4}(0.409)$ & \cellcolor{medium} $\mathbf{<1e-4}(-0.457)$ & \cellcolor{large} $\mathbf{<1e-4}(-0.577)$\\ \rowcolor{gray10}
            chatgpt & deepseek-coder & \cellcolor{small} $\mathbf{<1e-4}(0.329)$  & \cellcolor{small} $\mathbf{<1e-4}(-0.256)$ & \cellcolor{medium} $\mathbf{<1e-4}(-0.398)$\\
            chatgpt & gpt-4 & $\mathbf{0.014}(0.052)$  & $0.165(0.026)$ & $0.880(0.038)$\\ \rowcolor{gray10}            
            chatgpt & speechless-codellama & \cellcolor{medium} $\mathbf{<1e-4}(-0.36)$  & $\mathbf{<1e-4}(-0.137)$ & \cellcolor{small} $\mathbf{<1e-4}(-0.296)$\\            
            chatgpt & wizardcoder & \cellcolor{small} $\mathbf{<1e-4}(0.262)$  & \cellcolor{medium} $\mathbf{<1e-4}(-0.39)$ & \cellcolor{large} $\mathbf{<1e-4}(-0.68)$\\ \rowcolor{gray10}
            code-millenials & deepseek-coder & $0.086(0.004)$  & $\mathbf{<1e-4}(0.119)$ & $0.088(-0.057)$\\
            code-millenials & gpt-4 & \cellcolor{small} $\mathbf{<1e-4}(-0.288)$  & \cellcolor{medium} $\mathbf{<1e-4}(0.398)$ & \cellcolor{large} $\mathbf{<1e-4}(0.527)$\\ \rowcolor{gray10}
            code-millenials & speechless-codellama & \cellcolor{large} $\mathbf{<1e-4}(-0.637)$ & \cellcolor{small} $\mathbf{<1e-4}(0.287)$ & \cellcolor{small} $\mathbf{<1e-4}(0.247)$\\
            code-millenials & wizardcoder & $\mathbf{2e-4}(-0.117)$ & $0.107(-0.032)$ & \cellcolor{small} $\mathbf{<1e-4}(-257)$\\ \rowcolor{gray10}
            deepseek-coder & gpt-4 & \cellcolor{small} $\mathbf{<1e-4}(-0.218)$  & \cellcolor{small} $\mathbf{<1e-4}(0.198)$ & \cellcolor{medium} $\mathbf{<1e-4}(0.356)$\\
            deepseek-coder & speechless-codellama & \cellcolor{large} $\mathbf{<1e-4}(-0.548)$  & $\mathbf{0.002}(0.099)$ & $\mathbf{<1e-4}(0.125)$\\ \rowcolor{gray10}
            deepseek-coder & wizardcoder &  $\mathbf{0.024}(-0.086)$ & $\mathbf{0.005}(-0.041)$ & \cellcolor{small} $\mathbf{<1e-4}(-0.162)$\\
            gpt-4 & speechless-codellama & \cellcolor{medium} $\mathbf{<1e-4}(-0.362)$ & $\mathbf{2e-4}(-0.129)$ & \cellcolor{small} $\mathbf{<1e-4}(-0.333)$\\ \rowcolor{gray10}
            gpt-4 & wizardcoder & \cellcolor{small} $\mathbf{<1e-4}(0.168)$ & \cellcolor{medium} $\mathbf{<1e-4}(-0.335)$ & \cellcolor{large} $\mathbf{<1e-4}(-0.634)$\\
            speechless-codellama & wizardcoder & \cellcolor{large} $\mathbf{<1e-4}(0.529)$  & \cellcolor{small} $\mathbf{<1e-4}(-0.214)$ & \cellcolor{medium} $\mathbf{<1e-4}(-0.364)$\\
            \midrule
            & functional & \cellcolor{medium} $\mathbf{<1e-4}(-0.462)$ & \cellcolor{large} $\mathbf{<1e-4}(0.639)$ & $0.140(-0.068)$ \\ \rowcolor{gray10}           
            & chatgpt  &$\mathbf{<1e-4}(-0.056)$ & \cellcolor{large} $\mathbf{<1e-4}(0.550)$ & $\mathbf{<1e-4}(0.027)$\\
            & code-millenials & \cellcolor{large} $\mathbf{<1e-4}(-0.640)$ & \cellcolor{large} $\mathbf{<1e-4}(0.764)$ & $\mathbf{<1e-4}(0.204)$ \\ \rowcolor{gray10}
            & deepseek-coder   & \cellcolor{large} $\mathbf{<1e-4}(-0.529)$ & \cellcolor{large}  $\mathbf{<1e-4}(0.693)$ & $0.164(0.11)$ \\
            & gpt-4    & \cellcolor{small} $\mathbf{0.019}(-0.323)$& \cellcolor{large} $\mathbf{<1e-4}(0.482)$ & \cellcolor{small} $\mathbf{<1e-4}(-0.219)$\\ \rowcolor{gray10}
            & speechless-codellama  & $1.0(0.068)$& \cellcolor{large} $\mathbf{<1e-4}(0.587)$ & $0.677(0.086)$\\            
            \multirow{-7}{*}{canonical} & wizardcoder  & \cellcolor{large} $\mathbf{<1e-4}(-0.516)$ & \cellcolor{large} $\mathbf{<1e-4}(0.705)$ & \cellcolor{medium} $0.677(0.385)$ \\
            \midrule
            & functional & \cellcolor{large} $\mathbf{<1e-4}(-0.763)$ & \cellcolor{large} $\mathbf{<1e-4}(-0.541)$ & \cellcolor{large} $\mathbf{<1e-4}(-0.699)$ \\ \rowcolor{gray10}
            & chatgpt  & \cellcolor{medium} $\mathbf{<1e-4}(-0.454)$ & \cellcolor{medium} $\mathbf{<1e-4}(-0.368)$ & \cellcolor{medium} $\mathbf{<1e-4}(-0.436)$ \\
            & code-millenials  & \cellcolor{large} $\mathbf{<1e-4}(-0.839)$ & \cellcolor{medium} $\mathbf{<1e-4}(-0.417)$ & \cellcolor{large} $\mathbf{<1e-4}(-0.538)$ \\ \rowcolor{gray10}
            & deepseek-coder & \cellcolor{large} $\mathbf{<1e-4}(-0.646)$ & \cellcolor{medium} $\mathbf{<1e-4}(-0.336)$ & \cellcolor{medium} $\mathbf{<1e-4}(-0.363)$ \\    
            & gpt-4 & \cellcolor{large} $\mathbf{<1e-4}(-0.806)$ & \cellcolor{large} $\mathbf{<1e-4}(-0.607)$ & \cellcolor{large} $\mathbf{<1e-4}(-0.815)$ \\ \rowcolor{gray10}   
            & speechless-codellama  & $1.0(-0.071)$ & \cellcolor{small} $0.119(-0.149)$ & $0.150(-0.076)$ \\            
            \multirow{-7}{*}{expert} & wizardcoder  & \cellcolor{large} $\mathbf{<1e-4}(-0.801)$ & \cellcolor{medium} $\mathbf{<1e-4}(-0.385)$ & \cellcolor{small} $\mathbf{0.003}(-0.281)$ \\
            \midrule
            functional & few-shot & \cellcolor{small} $\mathbf{<1e-4}(-0.224)$ & $\mathbf{<1e-4}(-0.012)$ & $\mathbf{2e-4}(-0.075)$ \\ \rowcolor{gray10}
            functional & guideline & $\mathbf{<1e-4}(-0.027)$ & $\mathbf{0.03}(0.06)$ & \cellcolor{medium} $\mathbf{<1e-4}(-0.379)$ \\
            functional & keyword & $1.0(-0.025)$ & $1.0(-0.069)$ & $\mathbf{1.0}(-0.06)$ \\ \rowcolor{gray10}
            functional & platform & $1.0(0.014)$& $1.0(0.009)$ & $\mathbf{1.0}(-0.011)$ \\
            keyword & platform & $1.0(0.038)$ & $1.0(0.05)$ & $0.09(-0.034)$ \\ \rowcolor{gray10}
            guideline & keyword & $\mathbf{<1e-4}(0.002)$ & $\mathbf{0.004}(-0.123)$ & $\mathbf{<1e-4}(0.294)$  \\
            guideline & platform & $\mathbf{<1e-4}(0.037)$ & $\mathbf{0.006}(-0.07)$ & \cellcolor{small} $\mathbf{<1e-4}(0.272)$ \\ \rowcolor{gray10}
            few-shot & guideline & \cellcolor{small} $\mathbf{<1e-4}(0.172)$ & $0.01(0.07)$ & \cellcolor{small}$\mathbf{<1e-4}(-0.247)$  \\
            few-shot & keyword & \cellcolor{small} $\mathbf{<1e-4}(0.187)$ & $\mathbf{1.0}(-0.049)$ & $0.49(-0.007)$  \\ \rowcolor{gray10}
            few-shot & platform & \cellcolor{small} $\mathbf{<1e-4}(0.224)$ & $\mathbf{1.0}(0.006)$ & $0.308(-0.033)$  \\
            \bottomrule
            \multicolumn{5}{l}{\textit{P-values lower than $\alpha = 0.05$ are shown in \textbf{bold}. Effect sizes: \marktext{large}{ Large } - \marktext{medium}{ Medium } - \marktext{small}{ Small } - \marktext{white}{ Negligible } - \marktext{gray10}{ Negligible }}}
        \end{tabular}
    }
\vspace{-7mm}
\end{table}

\begin{tcolorbox}
\textbf{Answer to $\mathbf{RQ_0}$} -- Generated code exhibits significantly different levels of energy efficiency, both across LLMs and hardware platforms. The magnitude of these differences varies across hardware platforms. 
\speechless solutions are most efficient on the server rather than on PC and \rpi, where \codemillenials and \wizardcoder achieved the best results, respectively.
Solutions generated by \gpt are the least energy-efficient across all hardware platforms.
\end{tcolorbox}


\noindent \subsection{Human VS LLM-generated code ($\mathbf{RQ_1}$)}
\label{sec:results_rq1}

\noindent \textbf{Data Exploration}. 
As shown in \Cref{fig:energy_violinplot} and \Cref{tab:descriptive_statistics}, canonical solutions provided in the EvoEval benchmark suite tend to use less energy than the functional ones generated by the LLMs on both server ($5.105$ vs $5.785$ kJ) and RPi ($0.434$ vs $0.460$ kJ). In contrast, on the PC, the energy consumption of canonical solutions tends to be higher than that of functional solutions ($12.396$ vs $8.681$ kJ). 
Additionally, canonical solutions on the server tend to be the most efficient ones, even when compared to all LLM-generated functional solutions, except for \speechless (in which the generated solution is slightly more efficient). This result is reversed on PC, where  canonical solutions are the most inefficient ones. On the RPi, instead, functional solutions generated by \codemillenials and \deepseek are the most efficient ones. 

\noindent \textbf{Statistical Analysis}. The Shapiro-Wilk and the Q-Q plots show a normal distribution of the energy data of canonical solutions. The energy data of the functional solutions is not normally distributed. The ART ANOVA test 
suggests a significant difference between the groups on the server and the PC ($< 2.22e-16$), while a non-statistical significant difference on the RPi ($0.140$). \Cref{tab:p_values_effect_size} presents the p-values and effect sizes of the post-hoc analysis, which compares the energy consumption of the canonical and functional solutions produced by each LLM. On the server, the analysis reveals a statistically significant difference between the canonical and functional solutions generated by each LLM, with the exception of \speechless ($1.0$). 
On the PC, the test comparing the canonical and functional solutions groups for each LLM consistently shows statistical significance. We notice no significant difference on the RPi between canonical and functional solutions generated by \deepseek ($0.164$), \speechless ($0.677$), and \wizardcoder($0.677$). The difference is statistically significant in comparison with \codemillenials and \chatgpt.

The effect size is \textit{medium} on the server comparing canonical to functional solutions ($-0.462$). We have \textit{large} effect sizes when comparing canonical solutions to \codemillenials ($-0.640$), \deepseek ($-0.529$), and \wizardcoder ($-0.516$). This result confirms the higher energy efficiency of canonical solutions on the server. The effect sizes are \textit{medium} or \textit{negligible} for the other LLMs. On the PC,  canonical solutions are less efficient than the functional ones. Indeed, the effect sizes are \textit{large} in both canonical vs functional ($0.639$) and LLM-specific comparisons. For \gpt, the effect size is \textit{medium} ($0.382$). The effect sizes on the RPi are all either \textit{negligible} or \textit{small}, except for \wizardcoder, where it is \textit{medium} ($0.385$). 

\vspace{-1mm}
\begin{tcolorbox}
\textbf{Answer to $\mathbf{RQ_1}$} -- Human-developed solutions (canonical) tend to be more efficient than LLM-generated ones (functional) on the server, but they are less efficient on the PC. On the RPi, this difference is less (statistically) prominent. 
The type of LLM does not heavily influence the results of this comparison.
\end{tcolorbox}
\vspace{-3mm}

\subsection{Impact of prompt engineering ($\mathbf{RQ_2}$)}
\label{sec:results_rq2}

\noindent \textbf{Data Exploration}. 
On the server, functional solutions exhibit lower energy usage ($5.785$ kJ) than those generated with other prompts, except for the few-shot ones ($5.416$ kJ), as indicated in \Cref{tab:descriptive_statistics}. We also observe that code generated using the guideline ($5.833$ kJ) prompts consume more energy. Note that the standard deviation of the collected energy measures is significantly high for the guideline and few-shot prompts, suggesting that the hints we provided in the prompts for helping the LLMs generate greener software led to higher variation in characteristics of the generated source code. 

\begin{figure}[htbp]
\vspace{-3mm}
    \centering
    \includegraphics[width=\linewidth]{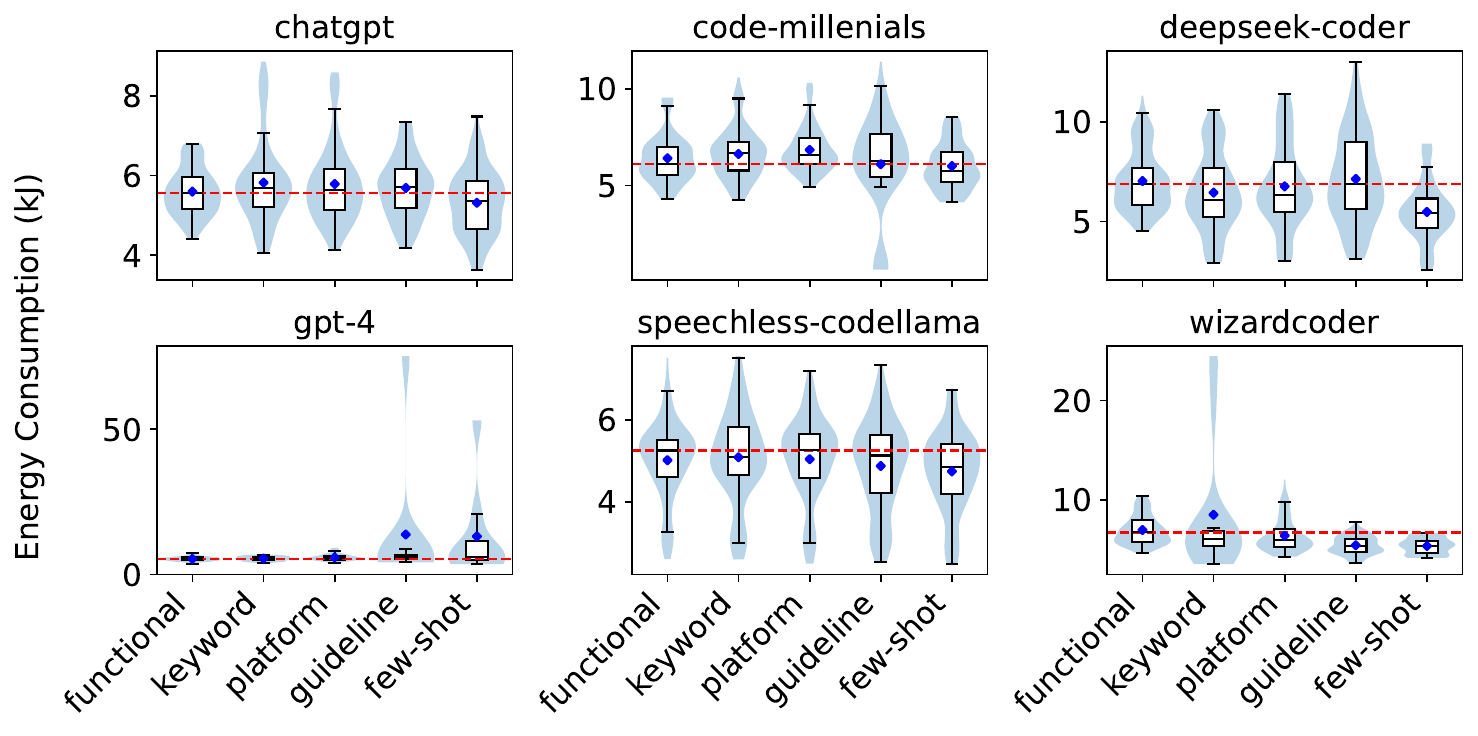}
    \vspace{-7mm}
    \caption{Energy consumption on the server of solutions generated via different prompting techniques.}
    \label{fig:multifaceted_violinplot}
\vspace{-2mm}
\end{figure}

This variation can be noticed in \Cref{fig:multifaceted_violinplot}, which shows how the energy usage varies across prompts for each LLM on the server. We see an unremarkable difference, except for \gpt, where a few data points influence the dispersion. We obtain similar violin plots on PC and RPi.
On the PC, the energy consumption for the code generated with the functional ($8.681$ kJ), keyword ($8.579$ kJ), and platform ($8.838$ kJ) prompts, and few-shot ($8.699$ kJ) is comparable, while the energy usage of guideline solutions is higher ($9.010$ kJ).
On the RPi, the solutions generated with the guideline prompt are the most energy-efficient ($0.407$ kJ), while the ones consuming the most are the functional solutions ($0.460$ kJ). The high dispersion observed in the server and PC is not present on the RPi. 

\noindent \textbf{Statistical Analysis}. 
The Shapiro-Wilk test (p-value $< 1e-4$) and the analysis of the Q-Q plots confirm that the data for $RQ_2$ does not follow a normal distribution. The ART ANOVA test evidences \textit{a statistically significant impact of the prompting strategies} on the energy efficiency of generated code. We also observe a significant interaction effect between the prompt type and the LLM used to generate the code. This result is consistent across hardware platforms. 

In the post-hoc analysis, we compare the combination of each prompting strategy. \Cref{tab:p_values_effect_size} include the p-values of the post-hoc analysis and the effect sizes of the Cliff's Delta test. The post-hoc analysis on the server shows a statistically significant difference between most of prompting strategies, except when comparing functional against keyword and platform ($1.0$) data, as well as keyword and platform ($1.0$).  The same result is seen on the PC, where, additionally, the comparison becomes also not statistically significant when contrasting few-shot and guideline solutions ($1.0$). On the RPi, the test yields a statistically significant difference between solutions generated via the functional prompt and all other prompts, confirming the observations done on the PC and server. The test results indicate no significant difference comparing the few-shot with keyword ($0.490$) and platform ($0.308$) solutions, as well as between keyword and  platform solutions ($0.09$). 

Despite the statistically significant differences observed, the effect size is small or negligible across all combinations of prompt engineering techniques across all hardware platforms (see \Cref{tab:p_values_effect_size}). 

\begin{tcolorbox}
\textbf{Answer to $\mathbf{RQ_2}$} -- Prompt engineering techniques tend to have limited impact on the energy consumption of LLM-generated Python code. We notice particularly higher variability in the energy usage of the solutions generated via the guideline and few-shot prompts on the server.
\end{tcolorbox}

\subsection{Green vs. LLM-generated code ($\mathbf{RQ_3}$)}
\label{sec:results_rq3}

\begin{figure}[htbp]
    \centering
    \includegraphics[width=\linewidth]{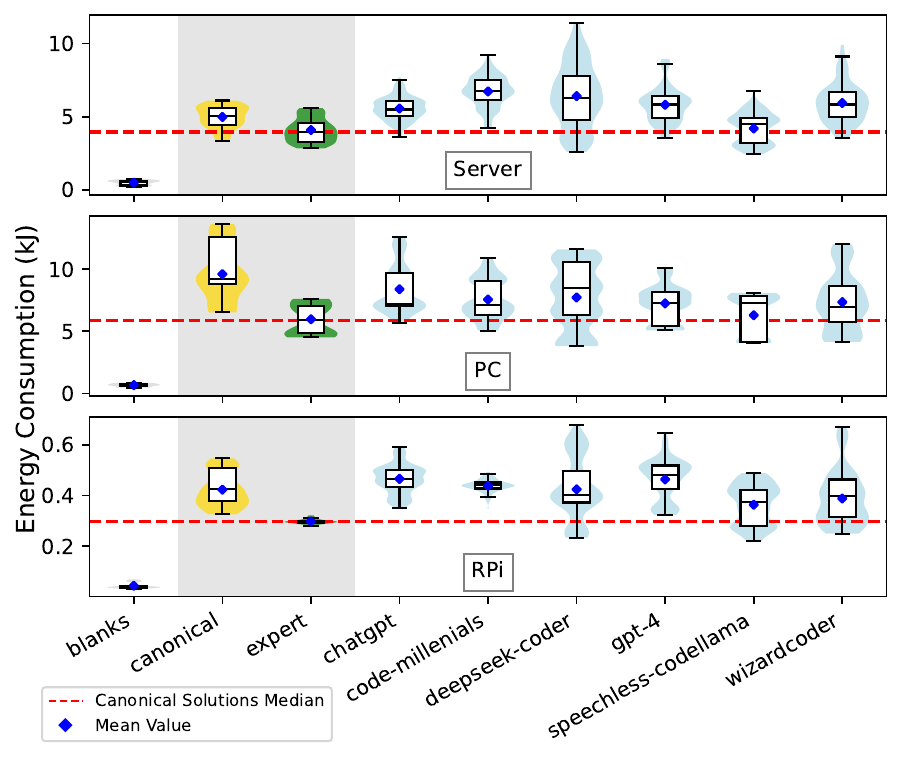}
    \vspace{-8mm}
    \caption{Energy of Green expert and LLM solutions.}
    \label{fig:energy_violinplot_rq3}
\vspace{-2mm}
\end{figure}

\noindent \textbf{Data Exploration}.
As shown in \Cref{fig:energy_violinplot_rq3}, expert-developed solutions tend to be consistently more efficient than \llm-generated ones across all hardware platforms. On the server the expert solutions are the most energy-efficient solutions ($3.962$ kJ), followed by \speechless solutions ($4.504$ kJ). The highest energy is used by \codemillenials solutions. On the PC, expert solution confirm to be the most energy efficient ($5.898$ kJ), followed by \wizardcoder solutions ($6.986$). The most inefficient are the solutions generated with \deepseek ($8.482$ kJ). Expert and \speechless solutions are the most energy efficient on the RPi, where their energy usage is $0.299$ kJ and $0.374$ kJ, respectively. We found some cases in which \speechless and \deepseek generate solutions that are more energy-efficient than expert code.  

\vspace{-3mm}

Analyzing each prompt-\llm pair on the server, we find that \speechless yields more energy-efficient solutions than the expert ($3.962$ kJ) under the few-shot ($3.926$ kJ) and guideline ($3.539$ kJ) prompts, where the latter solutions consume the lowest median energy usage on the server. The solutions generated from all other prompt-\llm pairs consume more energy than the solutions developed by the expert. The highest median energy is consumed by \codemillenials solutions generated with the keyword prompt ($6.811$ kJ).
The PC also showcases prompt-\llm combinations that yield more energy-efficient solutions than the ones developed by the expert ($5.898$), namely \wizardcoder via guideline ($5.713$ kJ) and few-shot ($5.870$ kJ) prompts, \speechless via the few-shot prompt ($5.441$ kJ). Overall, the most energy efficient solutions are those generated by \speechless with the guideline prompt ($5.441$ kJ), while the highest median energy is consumed by \codemillenials solutions generated using the few-shot prompt ($9.831$ kJ). 
On the \rpi, solutions generated by \wizardcoder ($0.297$ kJ) and \speechless ($0.283$ kJ) via the guideline  prompts are more efficient than the solutions developed by the expert ($0.299$ kJ). Overall, \speechless with the guideline prompt generates the most energy-efficient solutions ($0.283$ kJ), with the worst solutions generated by \gpt using the few-shot prompt ($0.518$ kJ).

\noindent \textbf{Statistical Analysis}. The Shapiro-Wilk test yields values lower (p-value < $1e-4$) than our significance level. The Q-Q plots confirm the results of the Shapiro-Wilk test. We use the ART ANOVA test to compare the groups. \Cref{tab:p_values_effect_size} includes the p-values of the ART ANOVA test and the effect sizes produced by the Cliff's Delta test to quantify the differences between the groups. ART ANOVA presents a \textit{significant difference} between expert code and all generated solutions on the server, PC, and RPi, except for \speechless, confirming the observations deriving from descriptive statistics. In \Cref{tab:p_values_effect_size}, the effect sizes are \textit{large} or \textit{medium} when comparing the expert code to the solutions generated by each LLM. This suggests that, in general, the expert code is more efficient. However, there are exceptions. The comparison between expert and \speechless solutions results in a \textit{negligible} effect on the server ($-0.071$) and RPi ($-0.076$), while it is \textit{small} ($-0.149$) on the PC. For \wizardcoder, the difference is \textit{small} on the RPi ($-0.281$).

\begin{tcolorbox}
\textbf{Answer to $\mathbf{RQ_3}$} -- The solutions developed by the Green software expert are more energy-efficient than almost all \llm-generated solutions across all hardware platforms, with large effect size. 
Some solutions generated by \speechless and \wizardcoder outperform those by the Green software expert. However, there is no consistent \llm-prompt pairing that consistently outperforms the solutions by the Green software expert across all hardware platforms.
\end{tcolorbox}

\section{Discussion}\label{s:discussion}


\subsection{Implications for developers}\label{sec:discussion_developers}

\punchline{Keep a critical attitude towards LLM-generated code}
The results of our experiment provide evidence that the energy efficiency of LLM-generated code is \textit{highly context-dependent}, varying considerably when targeting different hardware platforms (see \Cref{sec:results_rq0}) and the used LLMs (see \Cref{sec:results_rq1}). 
Our findings demonstrate that there is no single best LLM for generating energy-efficient code. Across all contexts, the execution environment plays a crucial role in determining actual performance. This insight is particularly important for practitioners, as LLM-generated code that appears efficient on one hardware platform might behave completely differently on another.
We encourage developers to consider at least the hardware platform where the generated code will be executed.
%
As a further step, we analyzed how each LLM performs with respect to individual coding problems and manually inspected the source code of solutions exhibiting low, medium, and high energy consumption (plots and raw data in the replication package~\cite{replicationpackage}). There, we observe the following: 
(i) \gpt is the only LLM able to generate code that uses external Python modules (\ie \texttt{NumPy} via the guideline and few-shot prompts), and in those cases, the solutions exhibited considerably higher energy consumption than the others probably because of the overhead from importing and initializing the \texttt{NumPy} library that is not amortized over the small computational workload as in this case; 
(ii) those solutions consuming higher amounts of energy primarily do so due to algorithmic inefficiencies (\eg nested \texttt{for} loops), rather than the usage of specific Python constructs like string concatenation, sorting, or slicing, which are known to have a higher theoretical complexity~\cite{python_complexity}; 
(iii) we observed the vast majority of outliers in the \rpi and much fewer in the server and PC, confirming that the hardware platform strongly interacts with the energy efficiency of the analyzed solutions.

    \punchline{Consider how LLM-generated code fits with the targeted Python environment and refine it accordingly}
    We know from the literature that a prominent role in energy efficiency is played by code-level characteristics, \eg in terms of used libraries~\cite{EASE_2024}, the presence of (anti) patterns~\cite{summerfield2013python}, and its interaction with the lower layers of the Python execution stack (\eg multithreading, compilation)~\cite{georgiou2020}.
    As such, we invite developers to review the LLM-generated code and manually refine it to enhance its efficiency, when needed. The guidelines we established in phase 2 of this study can serve as an initial checklist for identifying areas of improvement. 
    A ready-to-use checklist for Green Python code based on our guidelines is available in the replication package~\cite{replicationpackage}.

\punchline{Prompt engineering seems to do not pay off}
According to our results for $RQ_2$, prompt engineering techniques tend to have an inconsistent and limited impact on the energy efficiency of the generated Python code on all hardware platforms. 
This phenomenon is also generally confirmed in other recent independent studies, despite those studies were considering different coding solutions, prompts, hardware platforms, and statistical analyses~\cite{treude2025generativeaiempiricalsoftware,cappendijk2024generating,cursaru2024controlled}.
So, overall, we have convergent evidence that giving more freedom to the LLM via the prompt does not strongly contribute to the energy efficiency of LLM-generated code. Also, in our manual analysis of individual solutions, we also noticed that prompting might even lead to solutions that are \textit{less} energy efficient. In addition, our findings also point towards future possibilities where LLMs could be fine-tuned with hardware or energy awareness built in, leading to more context sensitive and energy efficient code generation.


\vspace{-2mm}
\subsection{Implications for LLM vendors}\label{sec:discussion_llm-vendors}

\punchline{LLMs should consider (energy) efficiency as a first-class metric} Currently, LLMs perform relatively well in benchmarks targeting the functional correctness~\cite{codeassistants}. However, the situation is different when considering \textit{non-functional aspects} like energy efficiency. Our results show that LLMs perform similarly to human-developed code ($RQ_1$ and $RQ_2$), but they are worse compared to code developed by a Green software expert ($RQ_3$). This result aligns with another recent study targeting performance~\cite{huang2024effibench}, where \gpt-generated solutions are on average 3.12x slower than human-developed solutions. This gap represents both an environmental challenge and an economic opportunity, as energy-efficient code can directly lead to reduced operational costs for large-scale LLM deployments. This is a problem for LLM vendors, as confirmed in a recent large survey with 547 participants from both academia and industry, where the most negative opinions about the code provided by AI assistants concern the alignment with non-functional requirements and security aspects~\cite{codeassistants}. While LLMs continue to evolve rapidly, our analysis across six diverse models with different architectures, sizes, and training approaches provides a robust foundation for understanding this persistent challenge.


\punchline{Greener code is a responsibility for everyone involved in LLMs development}
The fact that LLM-generated code is not consistently and significantly energy-efficient is not merely a technical issue; more importantly, it is a \textit{responsibility} for both LLM vendors and us, researchers. It is well-known that developers today extensively use LLMs for their coding tasks~\cite{codeassistants,so_ai}. If the generated code is inefficient, we risk having \textit{thousands or even millions of inefficient lines of code going into production every day}, thereby missing a significant opportunity to reduce the CO2 emissions of our software systems.
We urge LLM vendors such as OpenAI, Google, Anthropic, and DeepSeek to invest in better techniques for generating energy-efficient code. Promising techniques include fine-tuning, Retrieval Augmented Generation (RAG), or even smaller domain-specific models dedicated to green code generation. As organizations increasingly prioritize both cost optimization and energy metrics, LLMs that can generate more efficient code may find competitive advantages in enterprise markets.
The \se research community is already active in this direction (see Section \ref{sec:related}) and is more than willing to contribute in various ways (read below).

\vspace{-3mm}
\subsection{Implications for researchers}\label{sec:discussion_researchers}

\punchline{The hardware platform has a significant impact on code energy efficiency, yet it is often overlooked}
Our study does not show consistent trends across hardware platforms. As demonstrated by \Cref{tab:descriptive_statistics}, we are not able to indicate a specific LLM to obtain code with higher energy efficiency. 
We also notice this discrepancy when comparing the energy of the solutions generated across prompts. Functional solutions are the most efficient on the server, while those generated using the platform and guideline solutions present higher energy efficiency on PC and RPi. Looking at \Cref{fig:energy_violinplot} and \Cref{fig:energy_violinplot_rq3}, we can observe the difference in terms of energy usage of the canonical and expert solutions across hardware platforms. In particular, in both cases, the execution on the PC introduces higher energy consumption. We recommend researchers to \textit{consider the hardware platform as a factor} when testing the energy efficiency of generated code, as results may considerably vary across different platforms. This aspect is frequently overlooked in existing experiments on the energy consumption of LLM-generated code \cite{cursaru2024controlled, cappendijk2024generating, huang2024effibench}. Lower levels of the Python execution stack can affect energy efficiency. However, other factors also play a role, including the OS, the availability of computing resources, and the temperature of the testing environment \cite{ardito2019methodological}. These factors can confound experiments on software energy efficiency, specially when targeting resource-constrained devices. 

\punchline{Learning about Green software is still a valuable skill}
The results of $RQ_3$ demonstrate the higher energy efficiency of expert-developed code compared to generated one. The observation holds on all hardware platforms with some exceptions, such as the solutions generated with \speechless using the few-shot and guideline prompts. Therefore, in most cases \textit{having expertise in Green software remains the most effective approach for implementing energy-efficient code}. We highlight the need for further research on combining LLMs and prompting strategies to generate more energy-efficient code, as some generated solutions did outperform expert-developed code. We advise \textit{caution} when integrating generated code into larger software systems, as it may negatively impact non-functional requirements \cite{10.1145/3597503.3608128}, such as energy efficiency. 
This point is significantly important, considering growing evidence that shows generated code is blindly integrated by software developers, especially when the output meets their expectations \cite{10.1145/3586030}.

\punchline{Practitioners need a Green code base for benchmarking}
Well-known benchmarks for measuring the accuracy of LLMs have generic problems that are not tailored to specific code quality attributes. We highlight the need for a code base of examples of Green software to effectively benchmark the capabilities of LLMs in generating scientific software. For instance, EvoEval \cite{evoeval} contain problems such as string matching and sorting, which are generic and do not consider energy efficiency as a criterion. We advise researchers in AI for Software Engineering to develop a benchmark for investigating the capabilities of LLMs in generating energy-efficient code. Addressing this gap requires concerted efforts to develop and maintain representative Green code examples for different programming languages, along with prompting strategies proven to be designed for energy-efficient code generation.  Additionally, Green code bases can further help advance research in Green Software, in particular Green AI, by i) providing a reference dataset for performing more empirical studies, ii) allowing the development of models that can generate energy-efficient code, and iii) providing a platform for researchers and developers to promote best practices in Green software development. This can further lead to the development of leaderboards that allow researchers/developers to select LLMs by comparing the accuracy of the generated code and the energy efficiency of the generated code.

\vspace{-4mm}
\section{Related Work}\label{sec:related}

The energy-efficiency of LLM-generated code has been studied only recently. 
Cursaru \etal \cite{cursaru2024controlled} conducted an experiment on a Raspberry Pi, generating code for three programming problems in C++, JavaScript, and Python using Code Llama. The study revealed that while human-written code was generally more energy efficient, generated JavaScript outperformed human code. Moreover, prompting Code Llama for energy efficiency did not yield improvements, and varying temperature settings had no significant effect.
%
Cappendijk \etal \cite{cappendijk2024generating} compared the energy efficiency of Python code generated by Code Llama and DeepSeek Coder for three coding problems on a laptop. Notably, switching from \texttt{while} loops to \texttt{for} loops can reduce energy consumption by up to 60\%, although the code using \texttt{for} loops was not always correct.
Tuttle \etal \cite{10765804} examined the energy consumption and execution time of eight LeetCode problems solved by eight LLMs on a single server. Two are the main differences with respect to our study: (i) they share two solution types with us (\ie \textit{functional} and \textit{keyword} prompts), but do not include solutions related to hardware platform, guidelines, and few-shot learning, and (ii) they also introduce an additional prompt for the chain of thoughts strategy.
Tuttle \etal found that only 5.7\% of LLM-generated solutions are more efficient than human code, and LLMs are inconsistent at generating energy-efficient Python code.
Vartziotis \etal~\cite{LLM4CODE_2024} evaluated the energy consumption (and other performance-related metrics) of the Python code solving 6 coding problems from LeetCode, as generated by GitHub Copilot, OpenAI ChatGPT-3, and Amazon CodeWhisperer. Their study confirms that (i) LLM-generated code tends to be less sustainable than code developed by humans and (ii) applying prompt engineering techniques leads to slightly more efficient code.
Solovyeva \etal~\cite{FORGE_2024} evaluated the correctness and energy efficiency of 53 solutions of LeetCode problems in 3 programming languages (Python, Java, and C++), as generated by GitHub Copilot, OpenAI ChatGPT 4o, and OpenAI o1-mini. The solutions were executed on two laptops with different hardware configurations. Their results show that (i) Python solutions have the highest pass@1 accuracy, but it is still very low, (ii) the energy efficiency of human- and LLM-generated solutions tends to be similar for Python and Java, but not for C++, and (iii) LLM-generated solutions tend to be  machine-agnostic, showing strong  energy correlation across the two laptops.
Overall, for all studies mentioned above, the main differences and complementarities with our study are the following: 
(i) in our study we cover three \textit{complementary} hardware platforms (\ie server, laptop, and RPi), whereas all other studies focus on a single type of hardware platform (we recall that hardware aspects are a key factor of the energy efficiency of software~\cite{guldner2024development}, (ii) in our study we consider solutions that are developed by a Green software expert, whereas all human-developed solutions in the other studies are only the canonical ones, and (iii) in our study we also propose a systematically-elicited set of guidelines for developing energy-efficient Python code. 

\vspace{-2mm}
\section{Threats to Validity}
\label{sec:threats}


\noindent\textbf{External Validity}.
This study focuses exclusively on Python code, so its results may not generalize to other programming languages. 
Nevertheless, our experiment covers a heterogeneous set of coding problems, demonstrating that even Python code can be optimized for energy efficiency. 
Future work should extend the evaluation of LLM-generated code to multiple programming languages. 
The selected problems include recursion, string manipulation, numerical calculations, and other common coding patterns that ensure a broad coverage of real-world scenarios. 
Although our hardware platforms do not cover the entire range of possible architectures, we analyze three distinct platforms to provide insights into how energy efficiency varies between different hardware architectures.
For solutions developed by the Green software expert, we do not claim that they are fully representative of all sustainable coding practices among professionals; however, they serve as a practical upper bound with respect to what can be practically achieved when aiming to develop energy-efficient Python code. 
Further, the rapid evolution of LLMs presents a potential threat to the long-term generalizability of our findings, as newer models may exhibit different energy efficiency characteristics. However, we mitigate this threat through several design choices such as selection of six diverse \llms spanning different architectures, sizes, and training approaches, providing a broad foundation for generalization as well as using systematically derived  energy-efficiency guidelines that are problem-agnostic regardless of the specific LLM used. 


\noindent\textbf{Internal Validity}.
%
LLMs inherently introduce variability due to their probabilistic nature of generating. For the reliability of our findings, we evaluate multiple coding problems (each with multiple test cases) and multiple LLMs.
To control external influences such as concurrent processes, all experiments are conducted in a controlled environment where only the test workloads run, ensuring minimal influence/overhead due to system processes.
Potential inaccuracies in energy measurement tools are mitigated by executing (and repeating) multiple experimental runs, improving the reliability of the obtained measures. 
To address the significant fluctuations in energy readings at runtime, we (i) repeat each individual trial of the experiment 21 times and (ii) utilize the highest available sampling frequency for each measuring tool, \ie 5Hz for the server and PC, and 5000Hz for the Raspberry Pi, ensuring consistency in within-device comparisons. 
To mitigate the effects of CPU temperature variations due to continuous code execution, we set a cooldown period of 60 seconds after each run, which also reduces the risk of the Monsoon Power Monitor tool overheating and dropping data. 

\noindent\textbf{Construct Validity}.
In our study we use one measurement tool for each hardware platform, \ie a software tool for the server and PC, and a hardware power meter for the \rpi.
To mitigate this potential threat, the tools used in this study are all peer-reviewed (EnergiBridge~\cite{Sallou_EnergiBridge_Empowering_Software_2023, durán2024energyconsumptioncodesmall, roque2024unveilingenergyvampiresmethodology}), and industry-standard (Monsoon Power Monitor~\cite{en12010027, schulman2011phone, 10.1145/2380445.2380502, cursaru2024controlled}) tools that have been tried and tested throughout their lifespans in multiple peer-reviewed studies.
In our experiments, we employed few-shot prompting with generic examples rather than designing them to specific algorithmic or data structure problems. This choice may limit the representativeness of the evaluation, as more contextually relevant examples could lead to different outcomes. While this aspect can be further investigated, the use of generic examples reflects the perspective of non-experts in energy efficiency interacting with the LLM.

\noindent\textbf{Conclusion Validity}.
The results described in Section \ref{s:results} are supported by a coherent statistical analysis, which mitigates potential measurement interferences and identifies outliers (which have been removed by using a standard procedure -- see \Cref{sec:phase5} for further details). The Q-Q plots and the Shapiro-Wilk normality test indicate that our data does not follow a normal distribution, necessitating the use of a non-parametric test (ART ANOVA in our analysis). Each trial, defined by a unique combination of LLM, coding problem, prompt technique, and hardware platform has been repeated 21 times to increase the number of data points and statistical power, a number deemed sufficient for reliability in existing literature in the area of Green Software Engineering~\cite{guldner2024development}.
%

\vspace{-6mm}
\section{Conclusion and future work}\label{s:conclusion}
In this study we determine whether \llms can generate energy-efficient code. Specifically, we investigate the energy efficiency of code generated by six state-of-the-art \llms by prompting them to generate solutions for nine Python problems using five prompting styles. We compared the energy efficiency of the generated Python solutions against each other and against two levels of human-implemented solutions.
We conducted the measurements on three different hardware platforms to increase the robustness of our conclusions. Overall, we executed 363 solutions on each hardware platform for a total of $\approx$881h of sheer execution time and collected $\approx$4.6B data points, which are then statistically analyzed. 
Based on our results, we offer recommendations for LLM vendors and future research. We advise developers to consult Green software experts or follow the Green Python guidelines from this study, rather than relying solely on \llms for energy-efficient code.

\section*{Acknowledgments}
We would like to thank the Network Institute (Vrije Universiteit Amsterdam) for providing the Nebula platform for prompting the LLMs used in this study.  

\pagebreak

\bibliographystyle{ACM-Reference-Format}

\end{document}